\documentstyle [12pt]{article}
\renewcommand{\baselinestretch}{1.3}
\title{ Heavy Fermion Production and the Symmetry Breaking Sector of the
 Electroweak Interactions\thanks{This research was supported in part by
 the U.S.~Department of Energy.}
}
\author{Taekoon Lee
        \\
        \\
        Department of Physics \\
        Columbia University \\
        New York, NY  10027}

\begin{document}
\maketitle
\begin{abstract}
 We point out that heavy fermion production through the fusion of the
longitudinal gauge bosons might be relevant in probing the
strongly interacting symmetry breaking sector of the electroweak
interactions, by showing the dependence of the one loop amplititude
for ($ w^{+} w^{-} \rightarrow \overline{t} t $) on the symmetry
breaking mechanism. The one loop amplitude for   ($ w^{+} w^{-} \rightarrow
\overline{t} t $) is calculated for the standard model and  extended
technicolour theory. Techni-rho meson exchange is also briefly discussed.
 We find at $m_{t} = 150 ~GeV$ the cross section
 of  top pair production in $ e^{+} e^{-}$
collisions is comparable in order of magnitude
 to that of the longitudinal gauge boson scattering.
\end{abstract}
\def\thepage{CU-TP-589}
\thispagestyle{myheadings}
\newpage
\renewcommand{\baselinestretch}{2}
\pagenumbering{arabic}
\addtocounter{page}{0}

\newcommand{\be}{\begin{equation}}
\newcommand{\ee}{\end{equation}}
\newcommand{\bear}{\begin{eqnarray}}
\newcommand{\eear}{\end{eqnarray}}
\newcommand{\cl}{\mbox{$\cal L$}}
\newcommand{\cu}{\mbox{$\cal U$}}
\newcommand{\al}{\mbox{$\alpha$}}
\newcommand{\muom}{\mbox{$\frac{\mu}{m_{H}}$}}
\newcommand{\pitwo}{\mbox{$\pi^{2}$}}
\newcommand{\mutwo}{\mbox{$\mu^{2}$}}
\newcommand{\higgs}{\mbox{${m_{H}}^{2}$}}
\newcommand{\weak}{\mbox{$m_{W}^{2}$}}
\newcommand{\alb}{\mbox{$\bar{\alpha}$}}
\newcommand{\nub}{\mbox{$\bar{\nu}$}}
\newcommand{\tb}{\mbox{$\bar{t}$}}
\newcommand{\vtwo}{\mbox{$v^{2}$}}
\newcommand{\lsubt}{\mbox{$ {\cal L}_{t}$}}
\newcommand{\wl}{\mbox{$ W_{L}$}}
\newcommand{\albo}{\mbox{$\bar{\alpha_{1}}$}}
\newcommand{\albt}{\mbox{$\bar{\alpha_{2}}$}}
\newcommand{\qtwo}{\mbox{$q^{2}$}}

     1. It is well known that the symmetry breaking sector( SBS ) of the
 electroweak
interaction is far from being fully understood. We simply do not know what
 underlying dynamics
causes the spontaneous symmetry breaking that gives  masses to the weak gauge
bosons. Although there are many scenarios proposed for the SBS, they may be
broadly divided into two classes depending on their strength of interactions,
 namely,
 the weak scenario and  the strong scenario.

   Typically, in theories that belong to the  weak scenario, there are one or
a few
scalar particles with light masses ( roughly $ \leq 200 GeV $)
 and small coupling
constants  so that amplitudes may be calculated perturbatively.
The minimal Higgs model with one light elementary  Higgs boson and the
supersymmetric
models are typical examples of this scenario. Since the searches for the
Higgs boson have
already started, we expect this scenario to be throughly tested in the near
 future.

 The theories in the category of the strong scenario are characterized by the
abscence of particles with small masses. The minimal Higgs model with a heavy
Higgs boson
and  technicolour theories belong to this group. In  technicolour theories, a
 new
QCD-like gauge interactions appear at the TeV scale and  break chiral symmetry,
generating the Goldstone bosons necessary for the weak gauge boson masses.
 Observation of a heavy Higgs bosons or techi-rho mesons would be  direct
evidences for this
scenario. In the case that searches for  direct evidence for the strong
scenarios fail, it may   still be  possible to study the strongly
interacting SBS's using the method first
proposed by Dobado and Herrero and independently by Donoghue and Ramirez
\cite{Herrero,dono}.
 They noticed
that the scattering of the longitudinal weak gauge bosons( LWB ), which are
  essentially
the Goldstone bosons from the symmetry breaking, can be described, in anology
to
the $ \pi \pi $ scattering, by a low energy effective Lagrangian ( LEL ) with
 a few
unknown coefficients which, in turn, depend on the symmetry breaking
mechanism.
Then, the scattering amplitudes at low energies ( typically
  $ 0.4 ~TeV \leq \sqrt{s}
\leq 1.2 ~TeV $ ) are calculated through  chiral perturbation theory.

 This note is concerned with the strong scenario. The motivation of this
paper is the simple observation that heavy fermion production may be
relevant in probing the underlying dynamics of the SBS's because the
interactions between fermions and the LWB's are proportional to the fermion
 masses.
The heavy fermion we are considering here is the top quark which has quite a
heavy
expected mass of $150 ~GeV$. Thus there is a possibility that the interaction
between  top quark and the SBS is large enough that  low energy scattering
 processes
 involving top quark may reveal some information on the SBS. As a first step
 to explore
this possibility we calculate  the cross section
for the top pair production by LWB
fusion to the one  loop in chiral perturbation theory and compare it with
 that of
the LWB scattering to see its magnitude and sensitivity to the SBS.
Also computed is the effective vertex of the ordinary fermions and the LWB's
through
techi-rho meson exchanges. The background analysis which is absolutely
necessary for applicablity
of our work to actual experiment is not discussed here.
  This note is organized as follows.  In sec.2,
the LWB scattering is briefly reviewed and in sec.3, the amplitude
 for
 $( W_{L} W_{L} \rightarrow t ~\bar{t})  $ is calculated in  the minimal
 standard
 model and   extended technicolour theory. In sec.4, the effective interaction
of the ordinary quarks and the LWB's induced by the techni-rho meson exchanges
is derived and
in sec.5, the cross section for  $( e^{+} e^{-} \rightarrow \bar{\nu}~\nu
(W_{L}W_{L}) \rightarrow \bar{\nu}~\nu ~\bar{t}~t)$
 is calculated and compared with that of the LWB fusion.

   2. The scattering amplititudes of the LWB's at energies higher than the
mass of the
weak gauge bosons $ m_{W} $ can be easily calculated with the help of the
equivalence
theorem \cite{equi}.
 The theorem states that the S-matrix element of a process with the LWL's
 as the
external particles is equivalent, up to  $  O \left(\frac{m_{W}}{\sqrt{s}}
\right) $, to
that of the process with the external LWB's being replaced by the corresponding
 Goldstone bosons. It should be noted that the LWB's need not be the sole
 external particles.
The general proof of the theorem is given in Ref.4. According to the theorem,
 the LWB's
scattering can be described by the following Lagrangian,

\be
\cl = \frac{\vtwo}{2}\cl_{2} + \cl_{4} + {\cal O} (\partial^{6}),
\ee
 where
\bear
  \cl_{2} & = &\frac{1}{2} Tr \partial\mu U^{+} \partial\mu U , \nonumber \\
  \cl_{4} & = &\al_{1} ~\left( Tr \left(\partial\mu U^{+} \partial\mu U
                 \right)\right)^{2} +
                    \al_{2} ~Tr \left(\partial\mu U^{+}  \partial\nu  U
                  \right) Tr \left(\partial\mu U^{+} \partial\nu  U \right),
\label{chiralaction}
\eear
with  $ v = ( \sqrt{2} G_{F} )^{\frac{1}{2}} = 246 ~GeV$, and
\be
 U(x) = \exp\left(i \frac{w^{a}}{v} \sigma^{a}\right).  \label{goldstone}
\ee
Here, $w^{a}(x),a = 1,2,3$, are the Goldstone bosons corresponding  to the
LWB's and
 $\sigma^{a}$ are the Pauli matrices. While the leading term $\cl_{2}$ is model
 independent, the
$\al_{i}$ in $\cl_{4}$ are  quite sensitive to the underlying dynamics. As is
 well known,
the coefficients $~\al_{i}$ are  renormalized beyond the tree level, and the
one loop
renormalized coefficients were calculated
 for the minimal standard model and  scaled-up QCD \cite{Herrero,pipi,dawson1}.
They are given by \cite{dawson1},
\bear
\al_{1,SM}(\mu) & = &\frac{1}{4}\left[ \frac{\vtwo}{2\higgs} +
 \frac{1}{16 \pitwo}
                 (\frac{9\pi}{ 4 \sqrt{3}} -\frac{37}{9}) -
 \frac{1}{48\pitwo}\ln(\muom)
                  \right],     \nonumber  \\
\al_{2,SM}(\mu) & = &\frac{1}{4}\left[ - \frac{1}{ 16\pitwo}(\frac{2}{9})
 -\frac{2}{
                48\pitwo} \ln(\muom)\right],
\label{cosm}
\eear
for the minimal standard model, with $ m_{H}$ being the Higgs mass, and
\bear
\al_{1,TC}(\mu) & = &\frac{1}{4}\left[-0.011 - \frac{1}{48 \pitwo} \ln
\left(\frac{\mu(Gev)f_{\pi}}
                 {v} \right) \right],   \nonumber \\
\al_{2,TC}(\mu) & = &\frac{1}{4}\left[0.0046 - \frac{1}{24 \pitwo} \ln
\left(\frac{\mu(Gev)f_{\pi}}
                 {v} \right) \right]
\label{coqcd}
\eear
for  scaled-up QCD where $~\mu$ is the
renormalization scale comming from the one loop diagrams of $\cl_{2}$.
 We note that the coefficients $\al_{i,TC}$ were determined experimentally.
The amplitude for $ (w^{+} w^{-} \rightarrow w^{+} w^{-}) $ to  one
loop  in chiral perturbation is given by,
\bear
 T_{+-+-} & = & -\frac{u}{\vtwo}  +  \frac{4}{v^{4}} \left[ 2 \al_{1}(\mu)
\left(s^{2} + t^{2}\right)
                          + \al_{2}(\mu) \left( s^{2} + t^{2} + 2 u^{2}\right)
\right]
                               \nonumber \\
          &   &\mbox{} +\frac{1}{\left(4 \pi \vtwo\right)^{2}}\left[
- \frac{1}{12}\left(9  s^{2} + u^{2} - t^{2}\right) \ln\left(-\frac{s}{\mu^{2}}
                          \right) \right.     \nonumber \\
          &   &\left.\mbox{} -\frac{1}{12} \left(9 t^{2} + u^{2} - s^{2}\right)
                          \ln\left(-\frac{t}{\mu^{2}}\right)
-\frac{ u^{2}}{2} \ln
                          \left(-\frac{u}{\mu^2}\right) \right] ,
\label{chiral}
\eear
where $ s,t$ and $u$ are the Mandelstam variables and $\al_{i}(\mu)$ are
given in (\ref{cosm}),
(\ref{coqcd}).
 The cross section for (
$  e^{+} e^{-} \rightarrow \nub ~\nu (\wl \wl) \rightarrow \nub ~\nu \wl
\wl $)
 using (\ref{chiral}) and the effective
 W-approximation \cite{dawson2} is presented in Fig.6.

    3. In this section, the low energy one loop amplitude for $( w^{+} w^{-}
 \rightarrow \overline{t} ~t )$ is calculated in the minimal standard model
 and  the extended technicolour
theory. At this level, the amplitude is already sensitive to the symmetry
 breaking mechanism.

  Let us begin with the minimal standard model. The Higgs sector and
the Yukawa coupling
 of  top quark in the  standard model is,
\be
\cl_{(Higgs+top)} = \partial\mu \Phi^{+} \partial\mu \Phi - \frac{
 g^2 (m_{H}^{0})^{2}}
{8 ( m_{W}^{0})^{2}}\left( \Phi^{+}
\Phi - 2 \frac{(m_{W}^{0})^{2}}{g^2}\right)^{2} + \left(-\frac{
g m_{t}^{0}}{2 m_{W}^{0}} ~\tb_{L} \Phi_{2} t_{R} + h.c.\right)
\label{action}
\ee
where $g$ is the $ SU(2) $ gauge coupling and  $m_{t}^{0}, m_{W}^{0}$ and
$m_{H}^{0} $
 is the bare mass of the top  quark, gauge boson and Higgs respectively. Here,
\be
\Phi = \left(\begin{array}{c}
        w^{+}   \\
        \frac{1}{\sqrt{2}}(\rho + w^{0})
        \end{array} \right)
\ee
 is the $SU(2)_{L}$ doublet scalar field, with $ w^{\pm}, w^{0}$ being
the Goldstone
bosons giving mass to  the weak gauge bosons, and $\rho$ representing the
massive Higgs field.
 To simplify the  calculation we  neglect the diagrams whose loop is
composed of
fermion and   scalar propagators, which are suppressed at least by a
 factor of $ m_{t}/
\sqrt{s} $  relative to the  other loop diagrams.
Some diagrams of this kind are shown in Fig.1.  Then, the
only nontrivial diagrams we need to calculate are such that a single
 massive Higgs line is
connected to the external top quarks, so the calculation essentially reduces
 to the evaluation
of the off-shell Higgs decay into two LWB's. For this calculation, we
 closely follow
Marciano and Willenbrock \cite{marciano},
 who studied the on-shell Higgs decay into $\wl^{+} \wl^{-}$
to  one loop. We note that in our approximation there is
no mass renormalization of the top quark because the tadpole diagram,
which is the only
remaining source of the renormalization, is absorbed into the scalar self
energy in the
Marciano and Willenbrock scheme. From the diagrams in Fig.2, we can easily
 see that the
one loop amplitude
can be written as,
\be
\tilde{T}_{(w^{-}w^{+}\rightarrow t\bar{t}),SM} = \frac{-i g m_{t}}{2
 m_{W}}~\overline{u}(p_{1})
v(p_{2}) \cdot {\cal A} (\qtwo),
\label{amp1}
\ee
where $ u(p_{1}), v(p_{2}) $ are the spinors for the top and
$ ~\cal A (\qtwo)$ is given as the multiplication of the wave
function renormalization factor $z_{w}$ of  $ w^{\pm}$, the Higgs propagator
 $ G_{H}$ and the
vertex function $ \Gamma$, with $ q^{2}
= (q_{1} + q_{2} ) ^{2}$ and $q_{i}$ the momentum of $ w^{\pm}$,
\be
 {\cal A} (\qtwo) = \frac{m_{W}}{m_{W}^{0}} \cdot G_{H}(\qtwo)\cdot
\left(-i ~\Gamma(\qtwo)\right)
\cdot z_{w}.
\label{decay}
\ee
  From  Marciano and Willenbrock, we have
\be
\frac{m_{W}}{m_{W}^{0}} = \frac{1}{\sqrt{z_{w}}},\hspace{.5in}
z_{w} = 1 + \frac{g^2 \higgs}{16 \pi^{2} m_{W}^{2}} \left(- \frac{1}{8}\right).
\ee
A detailed calculation shows that
\bear
G_{H}(\qtwo) & = &\frac{-i}{\higgs}\left[ 1 + \frac{q^{2}}{\higgs} -
\frac{g^{2}\higgs}{
16 \pi^{2} m_{W}^{2}}\left(\frac{3}{16}\right)\left( 1+ 2 \ln
\left(\frac{- q^{2}}{\higgs}
\right)\right) \right. \nonumber \\
             &   &\left. \mbox{} +  \frac{g^{2} q^{2}}{16 \pi^{2} m_{W}^{2}}
\left(\frac{3}{16}\right)
-  \frac{g^{2}\qtwo}{16 \pi^{2} m_{W}^{2}}\left(\frac{3}{8}\right)\left( 1+
 2 \ln\left(\frac{- q^{2}}{\higgs}\right)\right) \right],
\eear
and
\bear
 -i ~\Gamma(\qtwo) & = &\frac{-i g ~\higgs}{2 m_{W}} \left[1 -
\frac{g^{2}\higgs}{16 \pi^{2}
 m_{W}^{2}}\left( -\frac{37}{16} + \frac{9 \pi}{8 \sqrt{3}} - \frac{3}{8}
\ln\left(\frac{-q^{2}}
{\higgs}\right)\right) \right. \nonumber  \\
                   &   &\left.\mbox{} -\frac{g^{2} q^{2}}{16 \pi^{2}
m_{W}^{2}} \left(\frac{1}{16}
\right) \left(  1- 2 \ln\left(\frac{- q^{2}}{\higgs}\right)\right) \right],
\eear
 to  $ O\left(\qtwo\right)$.
Substituting these results into (\ref{decay}), we get,
\bear
 {\cal A} (\qtwo) & = &-\frac{g}{2 m_{W}}\left[ 1+ \frac{g^{2} \higgs}{16
\pi^{2} m_{W}^{2}}\left( \frac{33}{16}
- \frac{9\pi}{
8 \sqrt{3}}\right) + \frac{\qtwo}{\higgs} \right.   \nonumber   \\
                  &   &\left.\mbox{} + \frac{g^2 \qtwo}{16 \pi^{2}
 m_{W}^{2}}\left( 2 -
\frac{9\pi}{8 \sqrt{3}} -\frac{1}{4} \ln\left(\frac{-\qtwo}{\higgs}
\right)\right) +
  O\left(q^{4}\right)
\right].
\eear
Here $m_{H}$ and $m_{W}$ are the  physical masses of the Higgs and the
 gauge bosons  defined
as  the poles of their respective propagators.

 For  extended technicolour theories \cite{extc}, we consider the
simplest case with a single $SU(2)_{L} \times U(1)_{Y}$ family of techniquarks
 $Q=( U,D )$ and ordinary quarks $ q = (t,b). $  We assume chiral symmetry
is a good symmetry of TC interactions and it breaks down to isospin symmetry.
The interaction between  ordinary fermions and  techniquarks at
the symmetry
breaking scale is point-like and given as
\be
\sum_{A} c_{A} ~j_{\mu}^{A} j_{\mu}^{A},
\label{R1}
\ee
with $c_{A} = g_{ETC}^{2}/m_{A}^{2} $ and
\be
j_{\mu}^{A} = f_{1}^{A} \overline{t_{L}} \gamma^{\mu} U_{L} +
f_{2}^{A} \overline{t_{R}} \gamma^{\mu} U_{R} + \cdots + h.c,
\label{R5}
\ee
where $f_{1}^{A},f_{2}^{A}$ are constants and the terms neglected are not
important in our calculation.
Expanding (\ref{R1}) with (\ref{R5}) and using the Fierz transformation, we can
write (\ref{R1}) as,
\be
       \sum_{A} c_{A} ~\left( 2 ~\left( f_{1}^{A} {f_{2}
^{A}}^{*} ~\overline{t_{L}}
t_{R} ~\overline{U_{R}} U_{L} + h.c \right) -   \frac{1}{2}
\left( |f_{1}^{A}|^{2} ~\overline{t_{L}}
              \gamma^{\mu} t_{L} +   |f_{2}^{A}|^{2} ~\overline{t_{R}}
             \gamma^{\mu} t_{R}\right) \overline{U}\gamma_{\mu} U +
 \cdots\right).
\label{R2}
\ee
Now the first term in (\ref{R2}) is the interaction which is
 subject to chiral perturbation. The second term is relevant for techni-rho
meson exchanges and it is discussed in next section.
To derive the effective interaction between  top quark and the Goldstone
 bosons we
are considering, we simply substitute $\overline{U}_{R}U_{L}$ in (\ref{R2})
with $
 <\overline{U}_{R}U_{L}> ~ U_{11}$, with $
< \overline{U}_{R}U_{L}>  $ being the techni-quark condensate, noting
that $ \overline{Q}_{R} Q_{L}$ and $ U$ defined in (\ref{goldstone})
 transform  identically under the chiral symmetry $ SU(2)_{L}\times
SU(2)_{R}$.  Then the LEL, at the tree level, is given by
\bear
\cl_{eff,TC} &=& \frac{\vtwo}{2} ~\cl_{2} + ~\cl_{t} +
\al_{TC} ~\cl_{t}\cl_{2},
 \nonumber \\
\cl_{t}      &=& - m_{t} ~\overline{t}_{L} ~t_{R} U_{11} + h.c,
\eear
with  $\al_{TC}= 0 $ and,
\be
m_{t} = - \left(\sum_{A} c_{A} ~Re\left(f_{1}^{A} {f_{2}^{A}}^{*}\right)\right)
\cdot < \overline{U}U >.
\label{R3}
\ee
Although $\al_{TC} =0 $ at the tree level, we will see that it can not be
zero at the one loop level because of  renormalization.
The amplitude to  one loop in $ \cl_{2}, \cl_{t}$ and to the tree level
in $ \cl_{2}\cl_{t} $ in the $\overline{MS}$ scheme is given by (see Fig.3)
\be
\frac{i m_{t}}{v^{2}} \overline{u}(p_{1}) v(p_{2}) \left[ 1 + \al_{TC}(\mu)
 q^{2} +
\frac{2 \qtwo}{16 \pi^2 v^2}\left( 1 - \frac{1}{2} \ln \left(\frac{-\qtwo}{
\mu^{2}}\right)\right)\right].
\label{loop}
\ee
 To find $\al_{TC}$ to one loop, we
follow the Weinberg
prescription \cite{weinberg} that requires the amplititude to be indepedent
 of the renormalization scale$~\mu$.
This gives,
\be
\al_{TC}(\mu) = \frac{1}{16 \pi^{2} \vtwo}\ln\left(
\frac{\mu_{0}^{2}}{\mu^{2}}\right),
\label{R6}
\ee
where $\mu_{0}$ is  a constant. Though the  exact value of the cut-off
$\mu_{0}$ should be determined by exprement,
 it is quite obvious that the scale
of $\mu_{0}$
 should be the symmetry breaking scale $4 \pi v,$
at which a new physics appears.
 Since no experimental data  for $\mu_{0}$ is  available, we simply take
$\mu_{0} = 4 \pi v $ in sec.5.
  Substituting (\ref{R6}) into (\ref{loop}), we have
\be
\tilde{T}_{(w^{-}w^{+}\rightarrow t\bar{t}),TC} =  \frac{i g^{2}m_{t}}{4
m_{W}^{2}} \overline{u}(p_{1}) v(p_{2})
\left[ 1  +  \frac{2\qtwo}{16 \pi^{2}\vtwo} \left( 1 - \frac{1}{2}
\ln\left(\frac{-\qtwo}{\mu_{0}
^{2}} \right)\right) \right].
\label{amp2}
\ee

   4. The techni-rho meson exchanges occur through the second term in
(\ref{R2}),
   \be
               -\frac{1}{2} \sum_{A} c_{A} \left( ~|f_{1}^{A}|^{2}
               ~\overline{
               t_{L}} \gamma^{\mu} t_{L} +   |f_{2}^{A}|^{2} ~\overline{
               t_{R}} \gamma^{\mu} t_{R}~\right) \overline{U}\gamma_{\mu} U.
   \ee
   Using the isospin symmetry in the TC sector and the
``current-field identity''  in the vector meson dominance hypothesis
\cite{bern},
   \be
   j_{\mu}^{a} = - \frac{m_{\rho}^{2}}{f_{\rho}} ~\rho_{\mu}^{a},
   \ee
   where $j_{\mu}^{a}$ is  the isospin current and $f_{\rho}$ is the
$\rho$-meson decay constant,
    we get  the vertex of  top quark and the techni-rho meson,
   \be
               \frac{1}{2} ~\frac{m_{\rho}^{2}}{f_{\rho}}
               ~\sum_{A} c_{A} \left( ~|f_{1}^{A}|^{2}
               ~\overline{
               t_{L}} \gamma^{\mu} t_{L} +   |f_{2}^{A}|^{2} ~\overline{
               t_{R}} \gamma^{\mu} t_{R}~\right)\rho_{\mu}^{0},
    \ee
    with the obvious notation that $m_{\rho}$,  $f_{\rho}$ now representing
the mass
    and the  decay constant of the techni-rho meson..
  Noting that
  \be
  \sum_{A} c_{A} \left( |f_{1}^{A}|^{2} + |f_{2}^{A}|^{2}
  \right) \geq  2 ~\sum_{A}  c_{A} |Re\left(f_{1}^{A} {f_{2}^{A}}^{*}\right)|
  \geq 2 \frac{m_{t}}{<\overline{U} U>},
  \ee
  we may write the general form of the interactions between
   ordinary quarks and  the techni-rho meson as
  \be
            \frac{m_{t}}{<\overline{U} U>} \frac{m_{\rho}^{2}}{
  f_{\rho}} \left( ~\left( C_{+}^{V} ~\overline{t}\gamma^{\mu} b  +
  C_{+}^{A} ~\overline{t}\gamma^{\mu}\gamma_{5} b \right) \rho_{\mu}^{+}
    +  \left( C_{0}^{V}~\overline{t}\gamma^{\mu} t +
 C_{0}^{A}~\overline{t}\gamma^{\mu}\gamma_{5} t \right)\rho_{\mu}^{0}
  + h.c ~\right),
  \ee
  where $ C_{a}^{V}, C_{a}^{A}$ are model-dependent, dimensionless constants
  of order one. Given a model, it should be straightforward to
 calculate  these constants. It  should also be noted that  there is no simple
  relations between these constants since the extended technicour interactions
  usually breaks the isospin symmetry. Now using the vertex between the
      rho mesons and  pions in QCD,
  \be
  f_{\rho} ~\varepsilon_{abc}~\rho_{\mu}^{a} ~\partial_{\mu} \pi^{b} \pi^{c},
  \ee
  the effective vertex between  ordinary quarks and the LWB's can be
  easily derived. For example, the effective vertex between $\overline{t},t$
 and
  $w^{+},w^{-}$
  is,
  \be
  - \frac{i m_{t}}{<\overline{U} U>} \left( C_{0}^{V}~\overline{t}
\gamma^{\mu} t +
  C_{0}^{A}~\overline{t}\gamma^{\mu}\gamma_{5} t\right)\left(
  \partial_{\mu}w^{+} w^{-} - \partial_{\mu}w^{-} w^{+} \right).
  \ee
  The vacuum condensate can be evaluated by scaling up the QCD result, which
gives
  \be
  <\overline{U} U> \approx 17 v^3,
  \ee
  with $  N_{TC}=3 $.

   5. Considering the heavy background and the smallness of the
 cross section
 for the LWB scattering in $e^{+}e^{-}$
collisions, it seems to be very difficult to probe the SBS using
LWB scattering in electron-positron collider.
 However, just to see the sensitivity of the process  $(w^{+} ~w^{-}
\rightarrow \overline{t} ~t)$ to the
symmetry breaking mechanism, we calculate
the cross section of top pair creation through  LWB fusion in $ e^{-} e^{+} $
collision shown in Fig.5 and compare it with that of  $w^{+} ~w^{-}$
 scattering. Adding
the trivial  term in Fig.4 to the   amplitude we have calculated in  section 3,
we get the scattering amplitude,
\bear
T_{(w^{-}w^{+}\rightarrow t\bar{t})} & = &\tilde{T}_{(w^{-}w^{+}\rightarrow
 t\bar{t})} + \nonumber \\
                                     &   &\hspace{.17in}
 \frac{1}{2}\left|U_{tb}\right|^{2}\left(\frac{i g m_{t}}{2 m_{W}
}\right)^{2}\overline{u}\left(p_{1}\right)\left(1-\gamma_{5}\right)
\frac{i}{\not\! p_{1}- \not\! {q_{2}}}\left(1+\gamma_{5}\right) v(p_{2}),
\label{amp}
\eear
with $U_{tb}$  the Kobayashi-Maskawa matrix element and
$ \tilde{T}_{(w^{-}w^{+}\rightarrow
t\bar{t})}$   given in (\ref{amp1}) and (\ref{amp2})
for the minimal standard model and the technicolour theory  respectively.
Here, for simplicity, the amplitude from the techni-rho meson exchanges is not
included. In the second term in
(\ref{amp}), the mass of  bottom quark  has been neglected. The cross
section for
$(e^{-}e^{+} \rightarrow \nu ~\nub(\wl \wl) \rightarrow
\nu ~\nub ~t ~\bar{t})$ can be
easily found using the effective W-approximation. In this approximation,
 the cross section
is given by \cite{Herrero},
\be
\frac{d\sigma}{d\hat{s}} =  \int_{0}^{1}d x_{1} ~\int_{0}^{1} d x_{2}
\hat{\sigma}\left(
w_{L}w_{L}\rightarrow t \bar{t}\right) ~\delta \left( \hat{s} - x_{1} x_{2} s
\right) F_{w_{L}}
(x_{1})
F_{w_{L}}(x_{2}),
\label{crosec}
\ee
where  $\sqrt{s}$ is the C.M energy of the $ e^{+},e^{-}$ and
\be
 F_{w_{L}}(x) = \frac{\al}{4\pi \sin^{2}\theta_{W}}\frac{1-x}{x}
\ee
is the distribution function of the gauge boson $ \wl^{\pm}$ and
$\hat{\sigma}$ is
the cross section for the subprocess $\wl^{-}\wl^{+}\rightarrow t~\bar{t}$ and
$\sqrt{\hat{s}} = M_{t \bar{t}}$.
Integration over $x_{1}$ and $x_{2}$ reduces (\ref{crosec}) to
\be
\frac{d\sigma}{d M_{t \bar{t}}} = \frac{-4}{\sqrt{\hat{s}}}
\left(\frac{\al}{4\pi
\sin^{2}\theta_{W}}\right)^{2}\left[ \left( 1+ \frac{\hat{s}}{s}\right)\ln
\left(\frac{\hat{s}}{s}\right) + 2 \left(  1- \frac{\hat{s}}{s}\right)\right]
\hat{\sigma}(\hat{s}).
\ee
 The cross sections for the minimal standard model and the technicolour theory
 at $\sqrt{s} = 2 ~TeV$  using (\ref{amp})  with $U_{tb}=1,$
are plotted in Fig.6.
The plots show that the cross section and the sensitivity to the
symmetry breaking mechanism of the top pair production are  smaller but
 comparable in order
of magnitude  at $m_{t} = 150 ~GeV$
 to those of the LWB scattering. We also notice that there is not much
difference between Fig.6(a) and 6(b), which means that the top production
is quite insensitive to the SBS  at this energy. This can be easily
understood when we compare (\ref{amp1}) and (\ref{amp2}).
 The difference between
(\ref{amp1}) and (\ref{amp2}) is approximately proportional
to $ q^{2}(M_{\bar{t}t}^
{2})$ and so it is small at low $ q^{2}.$  Now  the luminosity
of  $w^{+},w^{-}$ with large $q^{2}$ is much suppressed at low $\sqrt{s}$
to make the
top production relatively insensitive to  the SBS. However, the luminosity
of $w^{+},w^{-}$ with large $q^{2}$ increases as $\sqrt{s}$  becomes larger.
Therefore, the top production would be more sensitive to the SBS at higher
energies.

   Whether the top production is relevant in probing the SBS can be only said
    after the background has been studied. As is well known, there
   is a strong background in the LWB scattering and there would probably be
   as strong or stronger background in the case of the top production too.
   In this note we do not attempt to study the background and leave it
   as an open problem.

   Finally, we would like to comment on $ \overline{t}, b $ production in
 proton collider
   through the techni-rho meson exchanges.  Any possible observation of
$ \overline{t}, t $
   production in proton collider must be greatly obstructed by the strong
   background of the gluon-gluon fusion. However, $ \overline{t}, b $
   production does not have such  background, though it does have other
strong, but weaker than the g-g fusion, background such as
    W-g  fusion. One additional
   advantage of $\overline{t}, b $ production over
     $\overline{t}, t $   production is that
   there is no  $\overline{t}, b $ channel, in leading order, in the standard
   model except the trivial tree diagram, while  $\overline{t}, t $
   can be produced by  the Higgs exchange.
   Whether these facts will be of any use need to be seen.

 I am very grateful  to  N.H.Christ,  S.Dawson, A.H.Mueller and V.P.Nair
 for helpful discussions and encouragment. I also wish to thank the referee
for useful suggestions.

\newpage
\parindent 0.0in
\hspace{2.5in} Figure Captions

Fig.1: Examples of loop diagrams that are suppressed.

Fig.2: The leading Feynman diagrams for $(w^{+}w^{-} \rightarrow
\overline{t}t)$.
Thick solid lines and dotted lines represent the massive Higgs boson and the
Goldstone bosons respectively.

Fig.3: The leading Feynman diagrams for $(w^{+}w^{-} \rightarrow
\overline{t}t)$ in
 extended technicolour theory.

Fig.4: The blobs in dotted lines represent the one loop wave function
renormalization of
$w^{\pm}$.

Fig.5: LWB scattering and $\overline{t}t$ production through LWB fusion in
$e^{+}e^{-}$
collision.

Fig.6: $\frac{d\sigma}{dM_{\bar{t}t}}$ and $\frac{d\sigma}{dM_{w^{+}w^{-}}}$
for $\overline{t}t$ and $w^{+}w^{-}$ production from vector boson fusion in
 $e^{+}e^{-}$
collisions at $\sqrt{s}=2 ~TeV$. Dotted lines in (a),(b),(c) and (d)
represent the
tree level cross sections. (a),(b);$\overline{t}t$ production in the
extended technicolour
theory and the minimal standard model respectively.
 The solid and dot-dashed lines
in (b) are  at $m_{H} =
2 ~TeV, 1.2 ~TeV$ respectively.
(c),(d);$w^{+}w^{-}$ scattering in scaled-up QCD and the minimal standard
model respectively. The solid and dot-dashed lines in (d) are at
$m_{H} = 2 ~TeV,
1.5 ~TeV$ respectively.

%
%
%\end{description}

%
\end{document}